\documentclass[%
 reprint,
nofootinbib,
 amsmath,amssymb,
 aps,
 prd,
]{revtex4-2}

\usepackage{graphicx}
\usepackage{dcolumn}
\usepackage{bm}

\usepackage{tikz}
\usepackage{graphicx}
\usepackage{dcolumn}
\usetikzlibrary{snakes}
\usetikzlibrary{calc,decorations.markings}
\usepackage[compat=1.1.0]{tikz-feynman}
\usepackage{subcaption}
\usepackage{feynmp-auto}
\usepackage{hyperref}
\usepackage{nameref}
\usepackage{cleveref}
\usepackage{float}
\usepackage{physics}
\usepackage{cancel}
\usepackage{pgfplots}
\usepackage[normalem]{ulem}
\pgfplotsset{compat=newest}
\usepackage{tikz}
\usepackage{tikz-3dplot}
\usepackage{bigints}
\usetikzlibrary{intersections}
\usepackage{wasysym}
\usepackage{lipsum}
\usepackage[colorinlistoftodos]{todonotes}
\usetikzlibrary{3d,calc}
\usepackage{tikz-3dplot}
\Crefname{equation}{Eq.}{Eqs.}
\Crefname{figure}{Fig.}{Figs.}
\Crefname{definition}{Def.}{Defs.}
\Crefname{table}{Table }{Tables}

\newcommand{\BigO}[1]{\mathcal{O}( #1 )}
\newcommand{\dirac}[1]{\cancel{ #1 }}

\begin{document}

\preprint{APS/123-QED}

\title{Discrete Spacetime Theories Can Explain the Muon Magnetic Moment Discrepancy}

\thanks{The authors would like to thank Prof. Giovanni Amelino-Camelia for many fruitful discussions as we developed our idea.}%

\author{Paul C.W. Davies}
\email{paul.davies@asu.edu}
\author{Philip Tee}
\altaffiliation[Also at ]{p.tee@sussex.ac.uk, \\Department of Informatics, \\University of Sussex, Falmer, UK.}
\email{ptee2@asu.edu}
\affiliation{The Beyond Center for Fundamental Science,  Arizona State University, Tempe AZ 85287, USA}

\date{\today}

\begin{abstract}
An unsolved problem of particle physics is a discrepancy between the measured value of the muon anomalous magnetic moment and the theoretical prediction based on standard quantum electrodynamics.
In this paper we show that if spacetime possesses a fundamental length scale, the ensuing modifications to the photon propagator can account for the discrepancy if the scale is chosen to be $10^{-22}$~m; the corresponding energy being about $30$~TeV.
The possibility that spacetime possesses a graininess on a fine enough scale has a long history. 
One class of theories that develops this idea is Doubly Special Relativity (DSR), and we choose this as a model for our calculation. 
We note that the derived length scale is many orders of magnitude larger than the Planck length, but comparable to that of some higher dimensional gravitational theories. 
It is also within scope of experimental confirmation in the next generation of colliders.
\end{abstract}

\maketitle


\section{Introduction and Background}
\label{sec:introduction}

Fundamental physics is founded on the assumption of a spacetime continuum; for example, general relativity takes differentiable manifolds as a starting point. 
However, continuity is clearly an idealization as there will always be a finite precision to the measurement of space or time intervals. 
Furthermore, infinite divisibility goes hand in hand with unbounded frequencies, leading to the well-known problems of divergent quantities in quantum field theory. 
There is a widespread belief among theoretical physicists that spacetime must emerge from some form of small-scale substructure, or pre-geometry. 
Progress along this path has been hampered by the fact that no direct evidence for a substructure has been forthcoming from high-energy particle collisions, which probe space and time on intervals of $10^{-20}m$ and $10^{-28}s$ corresponding to a total energy of $14$~TeV \cite{evans2012large}. 
In spite of this, spacetime substructure may manifest itself indirectly via its effects on quantum field theory. In this paper, we examine one specific model of discrete spacetime and apply it to the problem of the magnetic moment of the muon, for which theory and experiment are somewhat discrepant.

The simplest model for spacetime discreteness is to introduce a fundamental length scale \cite{hossenfelder2013minimal,amelino2001testable}, thus dividing spacetime into identical cells. This ‘pixelation’ procedure may be regarded as a placeholder for a future, more sophisticated, treatment of spacetime emergence. 
Nevertheless, there may be observable consequences. 
The imposition of a fixed length obviously breaks local Lorentz invariance; bad news for quantum field theory (QFT). 
Fortunately, there are a number of workarounds. 
In so-called Doubly Special Relativity (DSR, which we use in this paper, Lorentz invariance is restored by introducing curvature into momentum space \cite{amelino2011principle}. 
An alternative approach involves non-commutative geometry \cite{adorno2011noncommutative,wang2002noncommutative,panigrahi2005induced}, which yields similar results.
It seems likely that there is a deep link between the two approaches. 
It is interesting to note that the theory of DSR has its origin in the work of Max Born in 1938 \cite{born1938suggestion} and Hartland Snyder \cite{snyder1947quantized} in 1947, and as such has almost as long a pedigree as QFT.

The pixel size is at this stage arbitrary. 
To avoid introducing an additional physical constant, it could be identified with the Planck length $l_P=\sqrt{\frac{\hbar G}{c^3}}=1.616\times10^{-35}m$, or some other scale that occurs naturally in existing theory. 
For example, some gravitational theories that invoke extra space dimensions derive an effective fundamental length scale many orders of magnitude greater than the Planck length \cite{hossenfelder2004minimal,lake2018does,anchordoqui2024large}. 
Here, we invert the problem by postulating that DSR accounts for the reported discrepancy in the muon magnetic moment \cite{logashenko2018anomalous,abi2021measurement,andreev2018final}, and use the experimental value of the discrepancy to determine the length scale. 
The result is a value consistent with existing high-energy particle physics data, but intriguingly close to both experimental limits and some non-standard gravitational theories. 
We predict a correction to the muon magnetic moment on account of the fact that DSR introduces dispersion into the free space propagators in Quantum Electrodynamics (QED). In what follows, we apply a photon propagator corrected to lowest order for pixilation effects to the standard calculation.

We start our analysis in \Cref{sec:propagators} with a brief overview of the computational framework, which we then apply in \Cref{sec:muon} to the computation of the anomalous magnetic moment.
The treatment of the computational framework is covered in more depth in \cite{davies2024quantum}; we relegate the details of our calculation to an Appendix.
We discuss the results and provide an outlook for further work in \Cref{sec:conclusion}.

\section{Computing the 1-loop Correction}
\label{computations}
\subsection{Modified propagators, Integration Measures and Vacuum Disperion}
\label{sec:propagators}

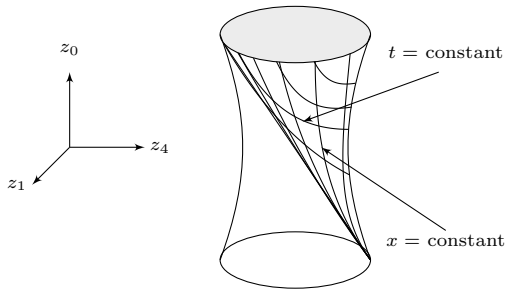
\begin{figure}[hbtp]
    \scriptsize
    \centering
    \begin{tikzpicture}[el/.style args={#1,#2}{draw,ellipse,minimum width=#1, minimum height=#2},outer sep=0pt,>=latex']
        \node(el-1) [el={2cm,0.75cm}] at (0,0){};

        \node(el-2) [el={2cm,0.75cm},fill=gray!15] at (0,3) {};
        \node at (-3,3.75) {$z_A z^A = z_0^2 - z_1^2 - z_2^2 - z_3^2 - z_4^2 = -\frac{1}{r^2}$};

        \draw (el-1.00) to[bend left=20] (el-2.00);
        \draw (el-1.-180) to[bend right=20] (el-2.-180);

        \draw[-] (-0.95,2.88) -- (1,0);

        \draw[thin, domain=0.0:1.0, samples=100, variable=\t]
            plot ({1 - 1.95*\t*exp(-0.1*\t)}, {2.76*(\t)});
        \draw[thin, domain=0.0:1.0, samples=100, variable=\t]
            plot ({1 - 2*\t*exp(-0.25*\t)}, {2.69*(\t)});
        \draw[thin, domain=0.0:1.0, samples=100, variable=\t]
            plot ({1 - 2*\t*exp(-0.5*\t)}, {2.64*(\t)});
        \draw[thin, domain=0.0:1.0, samples=100, variable=\t]
            plot ({1 - 2*\t*exp(-1.0*\t)}, {2.65*(\t)});
        \draw[thin, domain=0.0:1.0, samples=100, variable=\t]
            plot ({1 - 2*\t*exp(-2.0*\t)}, {2.75*(\t)});

        \draw[thin, domain=0.0:1.0, samples=100, variable=\x]
            plot ({-.95+1.67*\x}, {2.88*(1 - \x*exp(-0.5*\x))});
        \draw[thin, domain=0.0:1.0, samples=100, variable=\x]
            plot ({-.75+1.46*\x}, {2.75*(1 - \x*exp(-1*\x))});
        \draw[thin, domain=0.0:1.0, samples=100, variable=\x]
           plot ({-0.25+1.0*\x}, {2.64*(1 - 0.8*\x*exp(-1.25*\x))});
        \draw[thin, domain=0.0:1.0, samples=100, variable=\x]
           plot ({0.25+0.55*\x}, {2.64*(1 - 0.4*\x*exp(-1.3*\x))});

        \draw[->] (-3,1.5) -- (-3,2.5);
        \node at (-3,2.8) {$z_0$};
        \draw[->] (-3,1.5) -- (-2,1.5);
        \node at (-1.8,1.5) {$z_4$};
        \draw[->] (-3,1.5) -- (-3.5,1.0);
        \node at (-3.7,1) {$z_1$};

        \node [black] (1) at (0,1.84) {};
        \node [black,label={[align=left]$t=$ constant}] (2) at (2,2.5) {};
        \draw [->,black] (2.west) to  (1.east);

        \node [black] (3) at (0.25,1.5) {};
        \node [black,label={[align=left]$x=$ constant}] (4) at (2,0) {};
        \node [black] (5) at (2,0.3) {};
        \draw [->,black] (5.north) to  (3.east);



    \end{tikzpicture}  
    \caption{Embedding of a $3+1$ dimensional momentum space of constant curvature in a pseudo-Riemannian Minkowski space of $4+1$ dimensions. We control the embedding by setting the radius of the de Sitter hyperboloid as $r^2 \propto \eta^2$. An additional constraint is the choice of coordinates on the hyperboloid, that we make by choosing a flat slicing, using the parameterization in \Cref{eqn:flat1,eqn:flat2,eqn:flat3}. Overlaid are approximate curved lines of constant $x$ and $t$. One dimension is suppressed in this diagram.}
    \label{fig:desitter}
\end{figure}

In DSR the on-shell relation 
\begin{equation}\label{eqn:disp_stand}
    E^2=p^2+m^2 \text{,}
\end{equation} is modified by the addition of higher powers of $p$ on the right-hand side. 
The precise form of modification depends on the choice of momentum space curvature required to maintain local Lorentz invariance. 
There is no unique way to do this, but the coefficients of the powers of $p$ depend on the choice.
One `natural' approach is to use a de Sitter hyperboloid with positive curvature, embedded in a five-dimensional Minkowski-like momentum space (see \Cref{fig:desitter}). 
The coordinates, in analogy to Minkowski spacetime, have a timelike component and spacelike components, the former corresponding to energy, and the latter to momenta. 
Although de Sitter space has the virtue of naturalness and simplicity, there is still some ambiguity associated with the choice of slicing the de Sitter hyperboloid – a familiar issue when describing de Sitter space, for which there are three different slicings consistent with the symettries of the space that commend themselves (\cite{birrell1984quantum} \S 5.4).
If the slices are ‘horizontal’, describing closed sections, the coefficient of leading order correction turns out to be $0$. 
Thus, the lowest effective order is a quartic correction, as explored in \cite{davies2024quantum}.

There seems to be no compelling reason to eliminate the leading term correction for the generic case by this choice, so in what follows we adopt the more popular embedding, much studied in the DSR literature, of the so-called flat slicing of de Sitter momentum space \cite{amelino2012relative} (analogous to the steady-state theory of cosmology). 
To summarize the flat slicing construction, one chooses $P_0,\dots,P_4$ in a $5$ dimensional Minkowski manifold that constrain the $3+1$ momentum space coordinates $p_0,\dots,p_3$ to the hyperboloid $P_0^2-P_2^2-P_3^2-P_4^2-P_5^2=-\frac{1}{\eta^2}$, such that,
\begin{align}
    P_0 &= \frac{1}{\eta}\sinh \eta p_0 - \frac{\eta p_i^2}{2} e^{-\eta p_0} \text{,} \label{eqn:flat1}\\
    P_i &= p_i e^{-\eta p_0} \text{,}\label{eqn:flat2}\\
    P_4 &= -\frac{1}{\eta}\cosh \eta p_0 + \frac{\eta p_i^2}{2} e^{-\eta p_0} \text{,}\label{eqn:flat3}
\end{align}
where Latin indices denote spatial dimensions $i=1,2,3$ and Greek indices range $\mu=0,1,2,3$.
It is easy to see that this results in the metric,
\begin{equation}\label{eqn:metric_mom}
\begin{split}
    &g^{\mu\nu}=\begin{pmatrix}
        1 & 0 & 0 & 0\\
        0 & -e^{-2\eta p_0} & 0 & 0 \\
        0 & 0 & -e^{-2\eta p_0} & 0  \\
        0 & 0 & 0 & -e^{-2\eta p_0} \\
    \end{pmatrix}   \text{,~~}\\
    &g_{\mu\nu}=\begin{pmatrix}
        1 & 0 & 0 & 0\\
        0 & -e^{+2\eta p_0} & 0 & 0 \\
        0 & 0 & -e^{+2\eta p_0} & 0  \\
        0 & 0 & 0 & -e^{+2\eta p_0} \\
        \end{pmatrix}\text{.}
\end{split}
\end{equation}
This metric has a leading order metric determinant,
\begin{equation}
    \sqrt{-g} = 1 - 3 \eta p_0 \text{,}
\end{equation}
that must be used for momentum integrations to be covariant (as discussed in more detail in \cite{davies2024quantum}).
Associated with this metric, and following the treatment in \cite{amelino2012relative}, at leading order this results in the following dispersion relation for a massless particle,
\begin{equation}\label{eqn:cubic_dispersion}
    E^2=p^2 -  \eta p^3 \text{,}
\end{equation}
which changes the equation of motion to 
\begin{equation}
    \pdv[2]{\phi}{t} - \pdv[2]{\phi}{x} + i  \eta \pdv[3]{\phi}{x} = 0 \text{,} \\
\end{equation}
and produces a refractive index,
\begin{equation}
    n(p)=\frac{2\sqrt{1-\eta p}}{2-3\eta p} \label{eqn:refract_cubic}
\end{equation}
Here $p$ is the spatial momentum and $E$ the energy.
The coefficient $\eta$ has dimensions of inverse mass and is a surrogate for the fundamental length scale (in this work we have set $c = \hbar = G = 1$). 

For a particle of rest mass $m$ \Cref{eqn:cubic_dispersion} generalizes to the more complicated relation,
\begin{equation}
    m^2=E^2-\va{p}^2 + \eta E\va{p}^2 \text{,}
\end{equation}
\cite{amelino2012relative}.
However, in the experiment to measure the muon anomalous magnetic moment, the muons are highly relativistic. 
Therefore, to a good approximation the rest mass may be ignored, 
and \Cref{eqn:cubic_dispersion} used for the purposes of our calculation.

\subsection{Computing the Muon Magnetic Moment}
\label{sec:muon}

Quantum Electrodynamics (QED) predicts an anomalous gyromagnetic moment $g$ for charged fermions on account of vacuum polarization effects.
It was first calculated by Schwinger in 1948 \cite{schwinger1948quantum,schwinger1949quantum}, giving the famous result, $\frac{g}{2}=1+\frac{\alpha}{2\pi}$, where $\alpha \sim \frac{1}{137}$ is the fine structure constant.
This has subsequently extended to higher orders of perturbation theory, resulting in an expansion in terms of $\alpha$ that is in remarkable agreement with experiment: $0.28$ parts per billion for the electron \cite{hanneke2011cavity} - the equivalent of the width of a human hair to the distance between Los Angeles and New York.

This success is not as categorical in the case of the muon, for which its larger mass means that contributions from the electroweak and strong nuclear forces are confounding factors.
There is a persistent tension between the theoretical prediction and the measured value of $\frac{g}{2}$, that amounts to $(268 \pm 77) \times 10^{-11}$ at an uncertainty of $3.5 \sigma$ \cite{logashenko2018anomalous}.
In this work we will investigate under what circumstances this tension could be explained by the effect of a fundamental length scale altering the largest contributor to the calculation, the $1$-loop vertex function $\Gamma_\mu$ of QED for the muon.

To calculate the magnetic moment we evaluate the $1$-loop vertex function $\Gamma_\mu$ and considering the finite part of that calculation as it affects the Gordon relation (see for example chapter $9$ of \cite{ryder2009introduction}).
\vspace{0.1cm}
\begin{figure}[hbtp]
		\centering
        \begin{fmffile}{vertex}
        \begin{fmfgraph*}(100,175)
        \fmfleft{i1}
        \fmfright{i2}
        \fmftop{o1}
        \fmf{fermion,tension=1/3,label=$\normalsize \mbox{p}$,label.side=left}{i1,p1}
        \fmf{fermion,tension=1/3,label=$\normalsize \mbox{p-k}$,label.side=left}{p1,v1}
        \fmf{fermion,tension=1/3,label=$\normalsize \mbox{p'-k=p+q-k}$,label.side=left}{v1,p2}
        \fmf{photon,tension=0,label=$\normalsize \xrightarrow[\mbox{k}]{}$}{p1,p2}
        \fmf{fermion,tension=1/3,label=$\normalsize \mbox{p'=p+q}$,label.side=left}{p2,i2} 
        \fmf{photon,tension=1/3,label=$\normalsize \downarrow\mbox{q}$}{v1,o1}
            
        \end{fmfgraph*}
        \end{fmffile}
    \vspace{-2.5cm}
    \caption{Feynman diagram corresponding to the first order in $\alpha$ correction to the vertex function $\Gamma_\mu$, arising from QED interactions.}
    \label{fig:vertex}
\end{figure}
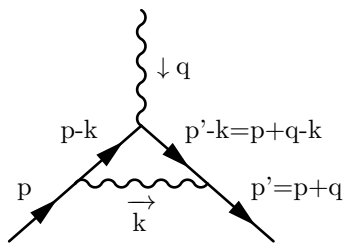

The details of the calculation are covered in \Cref{sec:appendix_p3} for the contribution from the modified photon propagator, and in \Cref{sec:appendix_muon} for the modified fermionic propagator.
In both cases we use the cubic modifications that arise from the modified dispersion relation of DSR, which change the conventional Feynman rules for the propagators as shown in \Cref{fig:feynman}.
\begin{figure}[hbtp]
        \raggedright
        \begin{fmffile}{frules}
        \begin{fmfgraph*}(55,40)
            \fmfleft{o1}
            \fmfright{o2}
            \fmf{photon, label=$k$}{o1,o2}
            \fmflabel{}{o1}
            \fmflabel{\Large $=\frac{1}{k_0^2-\va{k}^2(1- \eta \va{k}) + i\epsilon}$,}{o2}
        \end{fmfgraph*}\\
        \begin{fmfgraph*}(55,40)
            \fmfleft{o1}
            \fmfright{o2}
            \fmf{fermion, label=$p$}{o1,o2}
            \fmflabel{}{o1}
            \fmflabel{\Large $=i\frac{\dirac{p}+m}{p_0^2-\va{p}^2(1-\eta\va{p})-m^2}$,}{o2}
        \end{fmfgraph*}
        \end{fmffile}
        \caption{Modified Feynman rules for the photon and muon propagator. Note the introduction of the additional power of the three momentum.}
    \label{fig:feynman}
\end{figure}

As a result of the modification, the photon propagators alone violate Lorentz invariance. 
However, to calculate physical quantities one has to integrate over all momenta, and the introduction of momentum space curvature restores Lorentz invariance \cite{davies2024quantum}. 
The muon propagator, however, is not integrated over all momenta in the anomalous magnetic moment calculation, but is fixed to be “on shell.” 
By definition, the muon’s magnetic moment is defined in the muon’s rest frame. 
In conventional QED, that is unimportant because the theory is Lorentz invariant and so the result can be transformed to any frame. 
As a matter of fact, the current experimental measurement used muons at relativistic momenta in the laboratory frame: $3.094$ GeV/c (the so-called magic momentum) \cite{andreev2018final,abi2021measurement}. 
For the DSR calculation we need to fix a frame, and the muon’s rest frame $\va{p} = 0$ is the logical choice.

We summarize the results here for the $p^3$ photon and muon contribution,
\begin{equation}\label{eqn:p3_moment}
    \frac{g}{2} = 1 + \frac{\alpha}{2\pi} + \frac{\eta m_\mu \alpha}{3\pi}- \frac{8\eta \alpha^2}{3 \pi^{\frac{3}{2}}}\frac{p_\mu^3}{m_\mu^2} \text{,}
\end{equation}
In this equation $m_\mu=1.884\times10^{-28}kg$ is the mass of the muon, which is approximately $207$ time more massive than the electron.
In the rest frame of the muon the final term will vanish as $p_\mu=0$.
These effects will also be present for the electron, reduced by a corresponding factor of  $207$, consistent with the degree of agreement between experiment and theory for the electron.

\section{Experimental constraints on the length scale}
\label{sec:anomaly}

From the $1$-loop correction \Cref{eqn:p3_moment} we can use the experimental data to estimate the length scale coefficient $\eta$. 
This is shown in \Cref{tab:results}.

For the $p^3$ dispersion relation, we find that $\eta=1.84\times10^{22}kg^{-1}$, which corresponds to an energy scale of approximately $30.5$~TeV, still comfortably above the current crop of accelerator technology.
This is fourteen orders of magnitude smaller than the Planck scale.

We conclude that a $p^3$ dispersion modification from DSR will account for the anomalous magnetic moment discrepancy of the muon if the fundamental length scale is of the order of $10^{-22}m$.
The corresponding energy might be within the reach of the next generation of accelerators, making the theory testable in the not-too-distant future.
\begin{widetext}
\renewcommand{\arraystretch}{1.3}
\begin{table*}[hbtp]
    
    \setlength{\tabcolsep}{3pt} 
    \begin{center}
    \begin{tabular}{p{0.45\textwidth} p{0.2\textwidth} } 
        \hline\hline
        \textbf{Observable} & \textbf{$p^3$ Dispersion}  \\
        \hline
        Anomalous Moment $a_\mu$ ($\eta=\frac{1}{M_p}=4.60\times10^7kg^{-1}$) & $6.69\times10^{-24}$ \\
        Anomalous Moment $a_\mu$ ($\eta=1.84\times10^{22}kg^{-1}$) & $2.68\times10^{-9}$ \\
        \hline\hline
    \end{tabular}
    \caption{Summarized results at different values of $\eta$ for the anomalous magnetic moment.}\label{tab:results}%
    \end{center}
\end{table*}
\end{widetext}

\section{Discussion and future work}
\label{sec:conclusion}
We have shown that in a simple model (DSR), spacetime discreteness could account for the discrepancy of the muon magnetic moment. 
If this explanation is correct, it implies a correction to QED arising from quantum gravitational effects, since spacetime pixilation is being presented here as an idealized expression of some as-yet unspecified spacetime substructure. 
It is normally assumed that any quantum gravity effects in particle physics would be manifested only at the Planck scale of the order of $10^{16}$~TeV. However, in the absence of an accepted emergent spacetime formulation, the question of at what length or energy scale an underlying substructure might lead to observable consequences remains completely open. 
We note that alternative gravitational theories, such as the Randall–Sundrum models (RS) (also known as 5-dimensional warped geometry) \cite{randall1999large,randall1999alternative} involve an effective ‘Planck energy’ many orders of magnitude lower than $10^{16}$~TeV. 
Our result is also consistent with prior approaches using the apparatus of non-commutative geometry, where the experimental tension bounds the scale to be above $14$ TeV.

It is well known \cite{logashenko2018anomalous}, that the weak and strong forces will contribute a small correction to the muon’s anomalous magnetic moment, although a discrepancy remains. 
However, the anomalous magnetic moment of the muon has important contributions from both weak and strong nuclear forces. 
Although they are much smaller than QED effects, they account for some of the discrepancy between experiment and theory restricted to QED.

Although DSR is a popular theory that combines a fundamental unit of length with Lorentz invariance by introducing curvature into momentum space, it suffers from an arbitrariness concerning the choice of non-Euclidean geometry and coordinate slicing, necessary to proceed with any calculations. 
Unlike in general relativity, there is no principle of general covariance for curved momentum space, and the results will depend on the aforementioned choices. 
The use of de Sitter space sections commends itself on grounds of simplicity, but even so, the choice of coordinate slicing of the de Sitter hyperboloid will dramatically affect the results. 
The flat section slicing we use here we justify on the grounds that it leads to a non-zero lowest order correction to the dispersion relation.  
The closed section slicing, which suppresses the leading order, was used in our earlier work that addressed a different class of problems. 
In the absence of an additional principle, perhaps to emerge from a future complete theory, we are obliged to settle on a specific choice, on a problem by problem basis.

Taking our result at face value as a manifestation of quantum gravity effects in the muon magnetic moment encourages us to consider what other results in QED might provide an experimental cross-check for our hypothesis. 
Our approach could also be applied to both the electroweak and strong QED corrections. 
In addition, a direct confirmation of spacetime substructure at a fundamental length scale might be discovered by the next generation of colliders. 
\vspace{0.5cm}
\appendix
\section{Computation of 1-loop vertex correction with modified - \texorpdfstring{$p^3$}{p3} propagator}
\label{sec:appendix_p3}

The leading order anomalous magnetic moment arises from the 1-loop vertex function $\Gamma_\mu=\gamma_\mu+\Lambda_\mu(p,q,p')$, with the correction term calculated from the Feynman diagram \Cref{fig:vertex} \cite{ryder2009introduction}.
This was first calculated by Schwinger in 1948 \cite{schwinger1948quantum, schwinger1949quantum}, to first order with subsequent refinements improving the theoretical accuracy to $0.39$ parts per million.
The current value of the anomalous magnetic moment $a_\mu=116,591,821 (45) \times 10^{-11}$, includes corrections from both Electroweak and QCD interactions.
For a review of the current status of the calculations see Rich {\sl et al} \cite{rich1972current}.
The muon has additional corrections arising from Electroweak interactions \cite{logashenko2018anomalous}, but we will begin by computing the finite part of the vertex diagram, including a contribution from the modified propagator for electrodynamic only effects.
The effect of the fundamental length in spacetime is to modify the propagator for the virtual photon carrying four momentum $k$, by adding an additional term of $\eta \va{k}^3$ to the norm of the four momentum in the denominator.
Specifically we have,
\begin{equation*}
    k^2=k_\nu k^\nu=k_0^2 -\va{k}^2 + \eta \va{k}^3 \text{,}
\end{equation*}
that must be factored into the one loop vertex correction.

Using standard Feynman rules the diagram \Cref{fig:vertex} corresponds to the following integral, after performing the normal Feynman parameter combination of denominators,
\begin{widetext}
\begin{equation}\label{eqn:rawvertex}
    \Lambda_\mu(p,q,p')=\frac{2ie^2}{(2\pi)^4}\int\limits_0^1 \dd x \int\limits_0^{1-x} \dd y \int\limits_{-\infty}^\infty \dd^4 k  ~\frac{\gamma_\nu(\dirac{p}'-\dirac{k}+m)\gamma_\mu(\dirac{p}-\dirac{k}+m)\gamma^\nu}{[k^2-m^2(x+y)-2k(px+p'y)+p^2x+p'^2y]^3} \text{.}
\end{equation}
\end{widetext}
We could have combined the DSR correction using a third Feynman parameter, however this results in additional complications when finally evaluating the $x,y,z$ integrals.
In any case one should note, that the DSR correction involves the $3$-momentum $\va{k}$, and not the four vector $k$, and so it is only after we have performed the $k_0$ integration that we can approximate the denominator to obtain the leading order correction.

Our strategy relies upon the Gordon relation,
\begin{equation}\label{eqn:gordon}
\begin{split}
    \overline{u}(p') \gamma_\mu u(p) &= \frac{1}{2m}\overline{u}(p') \left [ (p_\mu+p_\mu') + i\sigma_{\mu\nu}q^\nu \right ]u(p) \text{, where} \\
    \sigma_{\mu\nu}&=-2i[\gamma_\mu,\gamma_\nu] \text{.}
\end{split}
\end{equation}
This permits us, once we have computed \Cref{eqn:rawvertex}, to express the term in $(p_\mu+p_\mu')$ as a coefficient of $\frac{i\sigma_{\mu\nu}q^\nu}{2m}$, from which corrections to $g$ (stated as corrections to $g/2$) can be read off \cite{dirac1928quantum}.
It will be recalled that the complete vertex operator $\Gamma_\mu(p,q,p')=(\gamma_\mu + \Lambda_\mu(p,q,p'))$, to first order, and to obtain the coefficient of $\sigma_{\mu\nu}$ the Gordon relation can be used to convert terms in $(p_\mu+p'_\mu)$ into terms in $\sigma_{\mu\nu}$.

Following the normal method we complete the square in \Cref{eqn:rawvertex}, by setting $k=k'+px+p'y$, noting that $\dd k'=\dd k$, and after substitution we switch the $k'$ back to a regular $k$ for cleanliness of notation.
This gives for our integral,
\begin{widetext}
\begin{equation}\label{eqn:squaredvertex}
    \Lambda_\mu(p,q,p')=\frac{2ie^2}{(2\pi)^4}\int\limits_0^1 \dd x \int\limits_0^{1-x} \dd y \int\limits_{-\infty}^\infty \dd^4 k  ~\frac{\gamma_\nu[\dirac{p}'(1-y)-\dirac{p}x-\dirac{k}+m]\gamma_\mu[\dirac{p}(1-x)-\dirac{p}'y-\dirac{k}+m]\gamma^\nu}{[k^2-m^2(x+y)+p^2x(1-x) +p'^2y(1-y)-2p.p'xy]^3} \text{.}
\end{equation}
\end{widetext}
We are only interested in the terms that produce the $(p_\mu+p_\mu')$ coefficients, and in any case those linear in $k$ will yield zero as the function becomes odd.
The terms quadratic in $k$ produce the divergent part of the vertex function which is dealt with by renormalization, although for our corrected interval they are partially regularized by the denominator.
This is to be expected as the length scale can be interpreted as a frequency cut-off.

After discarding $k$ terms our numerator becomes $N=\gamma_\nu[\dirac{p}'(1-y)-\dirac{p}x-\dirac{k}+m]\gamma_\mu[\dirac{p}(1-x)-\dirac{p}'y-\dirac{k}+m]\gamma^\nu$.
We need to simplify the expression by the commutation of the gamma matrices past the slashed four momenta noting that $\dirac{p}\gamma^\mu=2p_\mu-\gamma_]mu \dirac{p}$.
Further, to compute the $(p_\mu+p_\mu')$ coefficient we will be enclosing the expression in spinors and we have $\overline{u}(p')\dirac{p}'=m\overline{u}(p')$, and $\dirac{p}\overline{u}(p)=m\overline{u}(p)$, and so we can replace some of the slashed momenta by the fermion mass $m$.
After a laborious computation we have,
\begin{equation*}
    N=-4m(y-xy-x^2)p_\mu - 4m(x-xy-y^2)p_\mu' + F\gamma_\mu \text{\,}
\end{equation*}
where $F$ is an irrelevant complicated expression that does not contribute to our result.

We can now simplify the denominator by noting that $(p-p')^2=q^2=0$, and $p^2=p'^2=m^2$, which reduces it to $[k^2-m^2(x+y)^2]^3$.
When we include our modification to the propagator this in turn becomes $[k_0^2-\va{k}^2+\eta\va{k}^3-m^2(x+y)^2]^3$.
You will note that the sign of the correction is positive, which is consistent with a maximum velocity scale as the equation of motion produces propagation that is strictly slower than $c=1$.
From here we separate out the $k_0$ integral and convert to momentum space spherical polar coordinates (where we write $\va{k}^2=r^2$ and pick up a factor of $4\pi$ from the angular integrals) to obtain,
\begin{widetext}
\begin{equation}\label{eqn:transformed_p4}
    \Lambda_\mu(p,q,p')=\frac{8\pi ie^2}{(2\pi)^4}\int\limits_{0}^1\dd x\int\limits_{0}^{1-x}\dd y\int\limits_{-\infty}^\infty \dd k_0 \int\limits_{0}^\infty \frac{Nr^2}{[k_0^2-r^2+\eta r^3-m^2(x+y)^2]^3} ~\dd r \text{.}
\end{equation}
\end{widetext}
To make progress let us (for now) ignore the Feynman integrations and focus on the integrand (pulling a minus sign from the denominator),
\begin{equation}\label{eqn:p4_integrand}
    I=-\int\limits_{-\infty}^\infty \dd k_0 \int\limits_{0}^\infty \frac{Nr^2}{[r^2-\eta r^3+m^2(x+y)^2 -k_0^2]^3} ~\dd r \text{.}
\end{equation}
Writing $a^2=m^2(x+y)^2-k_0^2$ for economy of notation, we can expand the denominator around our Planck factor $\eta$, extracting a leading order correction as an additional term in the integrand as follows,
\begin{align}
    I_1&=-\int\limits_{-\infty}^\infty \dd k_0 \int\limits_{0}^\infty \dd r ~\frac{r^2}{(r^2+a^2)^3} \text{,}\\
    I_2&=-\int\limits_{-\infty}^\infty \dd k_0 \int\limits_{0}^\infty \dd r ~\ \frac{3\eta r^5}{(r^2+a^2)^4} \text{,}\\
    I&=I_1+ I_2\text{.} \label{eqn:integral_decomp}
\end{align}
These integrals can be evaluated directly as variants of the Euler Beta function,
\begin{equation*}
    B(x,y)=\frac{\Gamma(x)\Gamma(y)}{\Gamma(x+y)}-2\int\limits_0^\infty t^{2x-1} (1+t^2)^{-(x+y)} ~\dd t \text{,}
\end{equation*}
which upon substituting $x=\frac{1}{2}(1+\beta), y=\alpha-\frac{1}{2}(1+\beta)$ and $t=\frac{\varphi}{b}$, yields the following identify (see \cite{ryder2009introduction} \S 9A),
\begin{equation}\label{eqn:beta_func_ident}
    \int\limits_0^\infty \frac{\varphi^\beta}{(\varphi^2+b^2)^\alpha} ~\dd \varphi= \frac{\Gamma(\frac{1+\beta}{2})}{2(b^2)^{\alpha-\frac{(1+\beta)}{2}} \Gamma(\alpha)} \text{.}
\end{equation}
This formula is valid for $\Re x,\Re y > 0$, and integrals for other values of $x$ and $y$ are not finite.

The first integral in $r$ can be read off from \Cref{eqn:beta_func_ident} with $\beta=0$ and $\alpha=3$, resulting in the $k_0$ integral,
\begin{equation*}
    I_1=-\frac{\pi}{16}\int\limits_{-\infty}^\infty \frac{\dd k_0}{[m^2(x+y)^2-k_0^2])^\frac{3}{2}} \text{.}
\end{equation*}
If we multiply inside the denominator by $-1$, and bring the $-i$ outside, the resultant integral can be solved by substituting $k_0=m(x+y) \cosh \varphi$, and noting that for this even integral $\int\limits_{-\infty}^\infty\dd k_0=2\int\limits_0^\infty\dd k_0$, we arrive at the result, 
\begin{equation*}
    I_1=-\int\limits_{-\infty}^\infty \dd k_0 \int\limits_{0}^\infty r^2 \dd r \left \{ \frac{1}{(r^2+a^2)^3} \right \}  = -\frac{\pi i}{8} \frac{1}{m^2(x+y)^2}\text{.}
\end{equation*}
For $I_2$, we can use the Beta function identity \Cref{eqn:beta_func_ident} , which gives an integral of the form $\int\limits_0^\infty \frac{\dd x}{x^2-a^2}$ that can be evaluated to $\frac{\pi i}{2a}$ by the use of a contour integral around the pole at $x=a$.
This gives us for $I_2$ (after accounting for the sign reversing the $k_0^2$ and $m^2(x+y)^2$ terms in the denominator),
\begin{equation*}
    I_2=\eta\int\limits_{-\infty}^\infty \frac{\dd k_0}{k_0^2-m^2(x+y)^2} = \frac{\eta\pi i}{2m(x+y)} \text{.}
\end{equation*}

It remains to perform the resultant integral over the Feynman parameters.
Dealing with the two results for $I$ separately, we have for our first result,
\begin{widetext}
\begin{equation*}
    \Lambda^{(1)}_\mu(p,q,p')=-\frac{e^2}{4\pi^2m}\int\limits_{0}^1\dd x\int\limits_{0}^{1-x}\dd y ~\left \{ \frac{y-xy-x^2}{(x+y)^2} p_\mu + \frac{x-xy-y^2}{(x+y)^2} p'_\mu \right \}\text{.}
\end{equation*}
\end{widetext}
Each of the integrals over $x,y$ contribute a constant $\frac{1}{4}$, and recalling that the fine-structure constant $\alpha=\frac{e^2}{4\pi}$, and so we recover the standard result to first order in $\alpha$,
\begin{equation*}
    \Lambda^{(1)}_\mu(p,q,p')=-\frac{\alpha}{4\pi m}(p_\mu+p'_\mu) \text{.}
\end{equation*}

Turning to the second integral $I_2$, we have
\begin{widetext}
\begin{equation*}
    \Lambda_\mu(p,q,p')=\frac{\eta e^2}{\pi^2}\int\limits_0^1 \dd x \int\limits_0^{1-x} \dd y ~\left \{ \frac{(y-xy-x^2)p_\mu + (x-xy-y^2)p'_\mu}{(x+y)}\right \} \text{.}
\end{equation*}
\end{widetext}
Each term in the numerator when integrated against the denominator contributes $\frac{1}{12}$, and so we have our result,
\begin{equation*}
    I_2=\frac{\eta e^2}{12 \pi^2}(p_\mu + p'_\mu) \text{.}
\end{equation*}
Inserting into the Gordon relation and adding the $I_1$ result, for our $p^3$ propagator we have for our leading order magnetic moment,
\begin{equation*}
    \frac{g}{2} = 1 + \frac{\alpha}{2\pi} - \frac{2\eta m \alpha}{3\pi} \text{.}
\end{equation*}

So far our calculation is not manifestly Lorentz invariant as we have not taken into account the leading-order correction to the momentum curvature.
Using our result from \Cref{sec:propagators} this involved multiplying both the integrals $I_1$ and $I_2$ by $1-3\eta r$, noting that for a massless particle $p_0=\sqrt{p_1^2+p_2^2+p_3^2} = r$.
To $\BigO{\eta}$ this leaves us with one more integral to perform such that $I=I_1+I_2+I_3$ where,
\begin{equation}
    I_3=3\eta \int\limits_{-\infty}^\infty \dd k_0 \int\limits_0^\infty \dd r \frac{r^3}{(r^2 + a^2)^3} \text{.}
\end{equation}
This integral can be computed from the Beta function formula \Cref{eqn:beta_func_ident}, and the resultant integral in $k_0$ is elementary,
\begin{equation*}
    I_3=-\frac{3\eta}{2}\int\limits_0^\infty \frac{\dd k_0}{k_0^2 - m^2(x+y)^2} = -\frac{3\eta\pi i}{4} \frac{1}{m(x+y)} \text{.}
\end{equation*}
The resultant Feynman integrals are unmodified, but $I_2$,$I_3$ being of identical form do modify the coefficient from $\frac{\eta e^2}{\pi^2}$ to $-\frac{3\eta e^2}{2\pi^2}$.
This can be followed through the computation to give for our modification to the anomalous magnetic moment to $\BigO{\eta}$,
\begin{equation}
    \frac{g}{2} = 1 + \frac{\alpha}{2\pi} + \frac{\eta m \alpha}{3\pi} \text{.}
\end{equation}
It will be noted that including the Lorentz covariant measure the contribution to $\frac{g}{2}$ is now positive as expected.

\section{Computation of vertex contribution from modified - \texorpdfstring{$p^3$}{p3} Muon propagator}
\label{sec:appendix_muon}

In \Cref{sec:appendix_p3} we computed the correction to the vertex operator arising from the modified virtual photon propagator.
However, the vertex also includes two Fermion propagators, which although on-shell and so not subject to an integration over momenta, do contribute to the correction.

They appear in the calculation by virtue of the Dirac propagator,
\begin{equation*}
    S_F(p) = \frac{i}{\dirac{p} -m} = i\frac{\dirac{p}+m}{p^2-m^2} \text{.}
\end{equation*}
Modification of this propagator is complicated by the fact that unlike the photon these represent real on-shell fermions.
The momentum of the photon is ``integrated out'' in the calculation and is therefore not present in the result.
This is not possible when we deform the muon propagator.

Approaches to quantization of fermions in DSR have been considered, for example, Agostinit {\sl et al} and Harikumar {\sl et al} \cite{agostini2004dirac,harikumar2020quantization}, however a consistent set of Feynman rules is not available.
To make progress, we note that a key step in the evaluation of the vertex function $\Lambda_\mu(p,q,p')$ is the replacement of fermionic momentum norms such as $p^2$ by the rest mass $m^2$.
In the case of a massive particle however we have,
\begin{equation*}
    m^2=p^2+\eta \va{p}^3 {.}
\end{equation*}
When we make the substitution this will involve introducing a specific term for the momentum of the real, on-shell, fermion.
On first sight this appears an unsatisfactory outcome, however the details of how the anomalous momentum is measured \cite{abi2021measurement,andreev2018final} rely upon muons of a very specific momentum.
This value, the so-called ``magic'' momentum of $3.094$ GeV/c, allows the cancellation of terms in the classical electrodynamic interaction of the muons to simplify measurement.
We shall term this value $\va{p}_m = 3.094$ GeV/c, and so when we simplify our integral for the vertex function our substitution will be,
\begin{equation}
    p^2=m^2 - \eta \va{p}_m^3 \text{.}
\end{equation}
The computation of the contribution from the modified fermionic propagator, can now follow a very similar path to that for the virtual photon outlined in \Cref{sec:appendix_p3}.
We modify our fermionic propagator to,
\begin{equation}\label{eqn:muon_prop}
    S_F(p) = i\frac{\dirac{p}+m}{p^2 -\eta \va{p}_m^3-m^2} \text{.}
\end{equation}
Inserting this in our vertex function expression, simplifying and converting the photon momentum to polar coordinates changes \Cref{eqn:transformed_p4} to,
\begin{widetext}
\begin{equation}\label{eqn:transformed_muon}
    \Lambda_\mu(p,q,p')=\frac{8\pi ie^2}{(2\pi)^4}\int\limits_{0}^1\dd x\int\limits_{0}^{1-x}\dd y\int\limits_{-\infty}^\infty \dd k_0 \int\limits_{0}^\infty \frac{Nr^2}{[k_0^2-r^2+\eta (r^3 + f)-m^2(x+y)^2]^3} ~\dd r \text{,}
\end{equation}
where $f=\va{p}_m^3[(x+y)(x+y-1)]$, obtained in the simplification of the denominator of \Cref{eqn:squaredvertex} and captures the muon momentum dependence of the result. 
\end{widetext}

We proceed identically, expanding to leading order in $\eta$, and we recover the two integrals in \Cref{sec:appendix_p3}, but in addition, we have a further integration to consider,
\begin{equation*}
    I_m=-3\eta\int\limits_{-\infty}^\infty \dd k_0 \int\limits_0^\infty \dd r \frac{f r^2}{[r^2+m^2(x+y)^2 -k_0^2]^4} \text{,}
\end{equation*}
such that $I=I_1+I_2+I_3+I_m$ with reference to the decomposed integral \Cref{eqn:integral_decomp} in the prior section.

The momentum integral is readily performed using the Euler beta formula in the prior section and we obtain,
\begin{equation*}
    I_m=-\frac{\eta \sqrt{\pi}f}{8}\int\limits_{-\infty}^\infty \dd k_0 \frac{1}{[m^2(x+y)^2-k_0^2]^{\frac{5}{2}}} \text{.}
\end{equation*}
This integral can also be evaluated by making the substitution  $k_0=m(x+y) \cosh \varphi$, and reversing the sign of the denominator by factoring out a $(-1)^{\frac{5}{2}}=i$.
The result is,
\begin{equation}
    I_m=\frac{\eta i \sqrt{\pi f}}{6} \text{.}
\end{equation}
We can now perform the Feynman integrals, and we have (after expanding $f$),
\begin{widetext}
\begin{equation*}
    \Lambda_\mu(p,q,p')=\frac{\eta \sqrt{\pi}e^2}{3\pi^3}\frac{\va{p}_m^3}{m^3}\int\limits_0^1 \dd x \int\limits_0^{1-x} \dd y ~\left \{ \frac{(x+y)(x+y-1)[(y-xy-x^2)p_\mu + (x-xy-y^2)p'_\mu]}{(x+y)^3}\right \} \text{.}
\end{equation*}
\end{widetext}
The Feynman integral is not strictly convergent, as it leaves the following integral in $x$,
\begin{equation*}
    \int\limits_0^1 \dd x~ \left \{ \frac{3x}{2} -1 -\frac{1}{2x} -(1+x)\log x \right \} \text{,}
\end{equation*}
that has a value of $1$, plus an infinite part.
We can ignore the infinite part as being resolved in the renormalization process, and for the finite contribution to the magnetic moment for the fermion propagator modification we have,
\begin{equation}
    \frac{g}{2} = -\frac{8\eta \alpha^2}{3 \pi^{\frac{3}{2}}}\frac{\va{p}_m^3}{m^2} \text{.}
\end{equation}
We will discuss in the text , but the result is, as expected, dependent upon the value of the ``magic'' momentum $p_m$.
This momentum at $3.094$ GeV/c  is at an energy that is thirty times larger than the rest mass of the muon.
Overall, this would make a contribution nearly $30,000$ times larger than the photon propagator to the anomalous magnetic momentum, and would increase the energy scale needed to explain the tension in the measurement by four orders of magnitude.
It should also be noted that it operates in the opposite sense to the photon contribution.
This is unsurprising as the use of the physical propagator is not Lorentz conserving on its own, and there is no balancing factor from the momentum measure.
As discussed in the text, due to the absence of a fully Lorentz covariant treatment, we evaluate in the rest frame of the momentum where $\va{p}=0$.
\bibliography{PixelatedMuons}

\end{document}